\documentclass[final,authoryear,5p,times,twocolumn]{elsarticle}

\usepackage{amssymb}
\usepackage{epsfig}
 \usepackage{subfigure}
\usepackage{graphicx}
\usepackage{rotating}
\def\astrobj#1{#1}

\journal{New Astronomy}

\begin{document}

\begin{frontmatter}





\title{A Spectroscopic Study of \astrobj{DD UMa}: \astrobj{Ursa Major Group} Member and Candidate for BRITE}


\author[label1]{A. Elmasl\i\corref{cor}}
 \ead{elmasli@ankara.edu.tr}

\author[label2]{L. Fossati}
\ead{l.fossati@open.ac.uk}

\author[label3]{C. P. Folsom}
\ead{cpf@arm.ac.uk}

\author[label1]{B. Albayrak}
\ead{balbayrak@ankara.edu.tr}

\author[label4]{H.Izumiura}
\ead{izumiura@oao.nao.ac.jp}

\cortext[cor]{Corresponding author. Tel.: +90 312 212 67 20; fax +90
312 223 23 95}

\address[label1]{Ankara University, Astronomy and Space Sciences Department, Tando\u{g}an, 06100, Ankara, Turkey}
\address[label2]{Department of Physics and Astronomy, Open University, Walton Hall, Milton Keynes MK7 6AA, UK}
\address[label3]{Armagh Observatory, College Hill, Armagh, BT61 9DG, Northern Ireland}
\address[label4]{Okayama Astrophysical Observatory, NAOJ, NINS 3037-5 Honjo, Kamogata, Asakuchi, Okayama 719-0232, Japan}

\begin{abstract}

   {The \astrobj{Ursa Major group} is a nearby stellar supercluster which, while not gravitationally bound, is defined by co-moving members. \astrobj{DD UMa} is a $\delta$ Scuti star whose membership in the \astrobj{Ursa Major group} is unclear.

 The objective of this study is to confirm the membership of \astrobj{DD UMa} in the \astrobj{Ursa Major group}, as well as perform a detailed spectral analysis of the star.  Since \astrobj{DD UMa} is a low-amplitude $\delta$ Scuti star, we performed a frequency analysis.  We determined fundamental parameters, chemical abundances, and derive a mass and age for the star.

For this study we observed \astrobj{DD UMa} at the Okayama Astrophysical Observatory with the high-resolution spectrograph HIDES, between the $27^{th}$ of February and the $4^{th}$ March, 2009. Additional observations were extracted from the ELODIE archive in order to expand our abundance analysis. 
Group membership of \astrobj{DD UMa} was assessed by examining the velocity of the star in Galactic coordinates.  
Pulsational frequencies were determined by examining line profile variability in the HIDES spectra.
Stellar fundamental parameters and chemical abundances were derived by fitting synthetic spectra to both the HIDES and ELODIE observations. 

\astrobj{DD UMa} is found to be a member of the extended stream of the \astrobj{Ursa Major group}, based on the space motion of the star.  This is supported by the chemical abundances of the star being consistent with those of \astrobj{Ursa Major group} members.  The star is found to be chemically solar, with $T_{eff}=7450\pm150$ K and $\log g=3.98\pm0.2$.  We found pulsational frequencies of 9.4 c/d and 15.0 c/d.  While these frequencies are insufficient to perform an asteroseismic study, \astrobj{DD UMa} is a good bright star candidate for future study by the BRITE-constellation.}

\end{abstract}

\begin{keyword}

stars: $\delta$ Sct, -- stars: individual: \astrobj{DD UMa} 
stars: abundances -- technique: spectroscopic
\PACS 97.30.Dg \sep 97.10.Tk \sep 97.10.Sj \sep 98.65.Dxj




\end{keyword}

\end{frontmatter}


\section{Introduction}
\label{}

The \astrobj{Ursa Major group}, also known as the \astrobj{Sirius supercluster}, is the closest supercluster to the Sun. This group was first identified by the observation that five central stars in Ursa Major were moving towards the same point in space \citep{proctor, huggins}. After this discovery, a number of other stars were noticed to have similar space motions, some close to Ursa Major and some as far as the southern hemisphere (e.g HD 147584). This leads to distinguishing two subgroups in the \astrobj{Ursa Major group}: the nucleus (consisting of 13 stars including the five central stars in Ursa Major), and the extended stream. The subgroups are considered to share a common origin, but the members are considered gravitationally unbound. Stars located in the extended stream, as well as in the nucleus, provide information about the formation and evolution of the Ursa Major moving group. \astrobj{DD UMa} (18 UMa, $m_V$ = $4.^m832$) is a $\delta$ Scuti star located in the Ursa Major constellation, for which membership in the \astrobj{Ursa Major group} has not been well defined due to the lack of spectroscopic and photometric observations. 

Spectroscopic variability in \astrobj{DD UMa} was first noticed by \citet{schesinger}, based on observations from the Allengheny Observatory, who suggested the star was a spectroscopic binary candidate.  \citet{abt} obtained radial velocity measurements during 13 nights spread between December 1959 and April 1961, and reported that there was no variability in the radial velocity curve of \astrobj{DD UMa}, thus they concluded \astrobj{DD UMa} was not a spectroscopic binary.  \citet{percy} used  \astrobj{DD UMa} as a comparison star while photometrically observing HR 3775. In these observations they noticed that \astrobj{DD UMa} showed $\delta$ Scuti variability, with a $\sim$0.03 magnitude amplitude and a period of $\sim$3 hours.  \citet{horan} observed \astrobj{DD UMa} for two nights and confirmed the period and amplitude reported by \citet{percy}. The period derived by \citet{percy} is the only value available in the literature, and was recently included in work by \citet{rodriguez}, \citet{samus}, and \citet{watson}.

With this paper we supply a modern spectroscopic analysis of \astrobj{DD UMa}, which is notably absent in the literature.  
A detailed abundance analysis and frequency analysis of the star is important groundwork for further studies of the star, particularly asteroseismology, and helps place the star in a proper evolutionary context.  A careful assessment of \astrobj{DD UMa}'s membership in the Ursa Major moving group is also important. An older pulsating star such as \astrobj{DD UMa} could provide valuable information about the formation and evolution of \astrobj{Ursa Major group}.  Additionally, \astrobj{DD UMa} is a candidate target for the BRITE-Constellation\footnote{http://www.brite-constellation.at/}, which will perform high-precision photometry of bright stars in order to analyse their pulsational properties.  Thus a detailed ground based study of \astrobj{DD UMa} is valuable in advance of these potential space based observations.  

\section{Observations and Data Reduction}
We obtained a time-series of observations of \astrobj{DD UMa} between the $27^{th}$ of February and the $4^{th}$ March of 2009, with the high-resolution spectrograph HIDES (\textit{High Dispersion Echelle Spectrograph}) attached to the 1.88~m telescope at the Okayama Astrophysical Observatory. The data set covers a wavelength range of 3830-5600 \AA, at a resolving power of 47,000. An exposure time of 900 seconds was used for all observations. The observations from the $1^{st}$ of March (listed in Table 1) have a signal to noise ratio (S/N) of 95-420, however the other observations have a S/N lower than 50 due to poor weather conditions. Thus only the observations from the $1^{st}$ of March were suitable for a frequency and abundance analysis.

We reduced the time-series of HIDES observations of \astrobj{DD UMa} with standard IRAF (\textit{Image Reduction and Analysis Facility}) routines; this included bias subtraction, flat fielding, subtraction of background scattered light, extraction, and wavelength calibration using Th-Ar lamp spectra. After reduction and extraction, a barycentric correction was applied to the spectra, and the spectra were continuum normalised.  

Since hydrogen Balmer lines are sensitive to temperature and surface gravity, and insensitive to the details of chemical abundances, they are powerful tools for deriving stellar atmospheric parameters. 
While HIDES has high spectral resolution, the \'echelle spectral orders are only $\sim$85 \AA\, long.  The Balmer lines of \astrobj{DD UMa} are about 300 \AA\, wide, consequently proper normalisation of Balmer lines in the HIDES spectra was impossible.
Additionally, when performing abundance analysis, a data set with a longer wavelength range includes more spectral lines, which often provides access to abundances for additional elements. Because of this, we also used archival observations from the ELODIE instrument, which is a cross-dispersed \'echelle spectrograph that was attached to the 1.93~m telescope at the Observatoire de Haute-Provence.  These observations were extracted from the ELODIE archive\footnote{http://atlas.obs-hp.fr/elodie/} \citep{moultaka}, which provides reduced, optimally extracted, and wavelength calibrated spectra.  The ELODIE observations of \astrobj{DD UMa} cover a wavelength range of 3850-6800 \AA\, (R=42,000) and include Balmer lines suitable for our analysis. We used the highest S/N (=284) observation of \astrobj{DD UMa}, which was observed on the $28^{th}$ of March 2004. 


\begin{table}

\begin{tiny}

\begin{center}
\caption{The Julian date (HJD) and S/N of each observation obtained on the $1^{st}$ of March 2009 with HIDES.}

\begin{tabular}{cc|cc}
\hline\hline 
HJD & S/N  & HJD & S/N \\
\hline 
2454891.9845 & 	365  &  2454892.1202 & 	296\\
2454891.9577 & 	370  &  2454892.1410 & 	380\\
2454891.9687 & 	360  &  2454892.1519 & 	364\\
2454891.9838 & 	380  &  2454892.1629 & 	337\\
2454891.9947 & 	409  &  2454892.1738 & 	320\\
2454892.0058 & 	363  &  2454892.1849 & 	200\\
2454892.0168 & 	420  &  2454892.2135 & 	382\\
2454892.0545 & 	338  &  2454892.2245 & 	342\\
2454892.0833 & 	367  &  2454892.2394 & 	327\\
2454892.0943 & 	317  &  2454892.3658 & 	 95\\
2454892.1093 & 	395  &               &     \\
\hline
\end{tabular} 
\end{center}
\end{tiny}
\label{Table0}
\end{table}


\section{Frequency Analysis}
The frequency analysis of pulsating stars provides information about the interiors of those stars. The absorption lines obtained from high-resolution and high S/N time-series spectra of pulsating stars show line profile variations, which carry indications of a star's pulsational behaviour.  If these line profile variations can be resolved, they can be used for mode identification of a star's pulsation frequencies.

We performed frequency analysis with FAMIAS (\textit{Frequency Analysis and Mode Identification for Asteroseismology}, \citealt{zima8}) on the blended absorption lines centered at 4417, 4443, 4384, and 4481 \AA\, in the HIDES data set from the $1^{st}$ of March 2009. FAMIAS searches for periodicities in time-series spectra using a Fourier analysis.  These frequencies can then be refined using multi-periodic least-squares fitting. The program can also spectroscopically identify the pulsation modes associated with pulsation frequencies using either line moments \citep{briquet} or a Fourier parameter fit \citep{zima6}. 
When searching for pulsation frequencies we used the `pixel-by-pixel' mode, which computes Fourier spectra for each pixel in the line profile, then averages them together to produce a final mean Fourier spectrum.  

An initial frequency analysis was performed on the lines centered at 4417, 4443, 4384, and 4481 \AA, yielding peak frequencies between 9.4-10.8 c/d. For peaks in the Fourier spectrum, S/N is the ratio of the amplitude of the peak to the mean amplitude of the nearby spectrum after it has been pre-whitened with the peak's frequency. \citet{breger} found that a S/N amplitude ratio of 4 is a good criteria for distinguishing pulsation frequencies from noise. The only line that showed a frequency peak substantially above a S/N value of 4 during our frequency analysis was the line at 4384 \AA. Thus we chose to focus our frequency analysis on this absorption line, which is a blend of the Fe~I 4383.544 \AA\, and Fe~II 4385.387 \AA\, lines. 

The analysis of the  4384 \AA\, line yielded a main frequency at $f_1$=9.4 c/d (mean amplitude across the line A=0.0060, S/N=11.3) and another frequency at $f_2$=15.0 c/d (A=0.0043, S/N=6.2). The main frequency of \astrobj{DD UMa} at 9.4 c/d corresponds to a pulsation period of 2.6 hours (0.11 days). Unfortunately, the precision of these frequencies is limited due to the short time span of the usable spectra and the low pulsation amplitudes. Since the pulsation amplitudes of \astrobj{DD UMa} are very low, our time-series data set does not have sufficient S/N to resolve the details of the line profile variations, thus we could not perform mode identification on the obtained pulsation frequencies.

\section{Atmospheric Parameters and Abundance Analysis}
\label{AbundanceAnalysis}

For the spectral analysis we computed synthetic LTE spectra with the Synth3 code \citep{kochukhov}.  As input, we calculated plane-parallel, LTE model atmospheres with LLmodels \citep{shulyak}, which includes detailed line blanketing, and convection with the method of \citet{canuto}. Atomic line data were extracted from the Vienna Atomic Line Database (VALD, \citealt{piskunov,kupka,ryabchikova}).  

We first determined the effective temperature ($T_{eff}$) and surface gravity ($\log g$) making use of photometric calibrations applied to Str\"omgren and Geneva colors, obtaining average values of $T_{eff}$=7853 K and $\log g$=4.10. This set of parameters did not fit the H$\alpha$ line profile, therefore, we refined $T_{eff}$ by directly fitting the observed hydrogen line wings, obtaining $T_{eff}=7450\pm150$ K (see Fig. 1). We also examined the Fe I/Fe II ionisation equilibrium and, by enforcing this, derived a $\log g$ value of 3.85 from the ELODIE observation. The average of the photometric calibration and spectroscopic $\log g$ values is 3.98. The chemical abundances derived with this $\log g$ value, presented in Fig. 2, are in good agreement for different ionisation stages of the same element. For comparison, we calculated $\log g$ based on the mass derived in Sect. \ref{Membership-Evolutionary}, and the radius derived from the bolometric magnitude and effective temperature. We obtained $\log g$ = 4.06, which is consistent with our average $\log g$. The final fundamental parameters of \astrobj{DD UMa} are $T_{eff}=7450\pm150$ K and $\log g=3.98\pm0.2$.

We derived a microturbulence velocity ($\xi$)$=2\pm1$ km/s by minimising the standard deviation of weak and strong lines of both Fe I and Fe II, in the same way as described by \citet{fossati7}. We also calculated $\xi$=2.6 km/s from the relation given by \citet{pace}:
$\xi=-4.7\log(T_{eff})+20.9$ km/s. 
Since this value agrees well with the one obtained from the minimisation of standard deviations, we used $\xi=2.6\pm1.0$ km/s during our analysis.


\begin{figure}[htb]
\begin{center}  
 \includegraphics[trim = 0mm 0mm 0mm 0mm, clip, width=7.8cm]{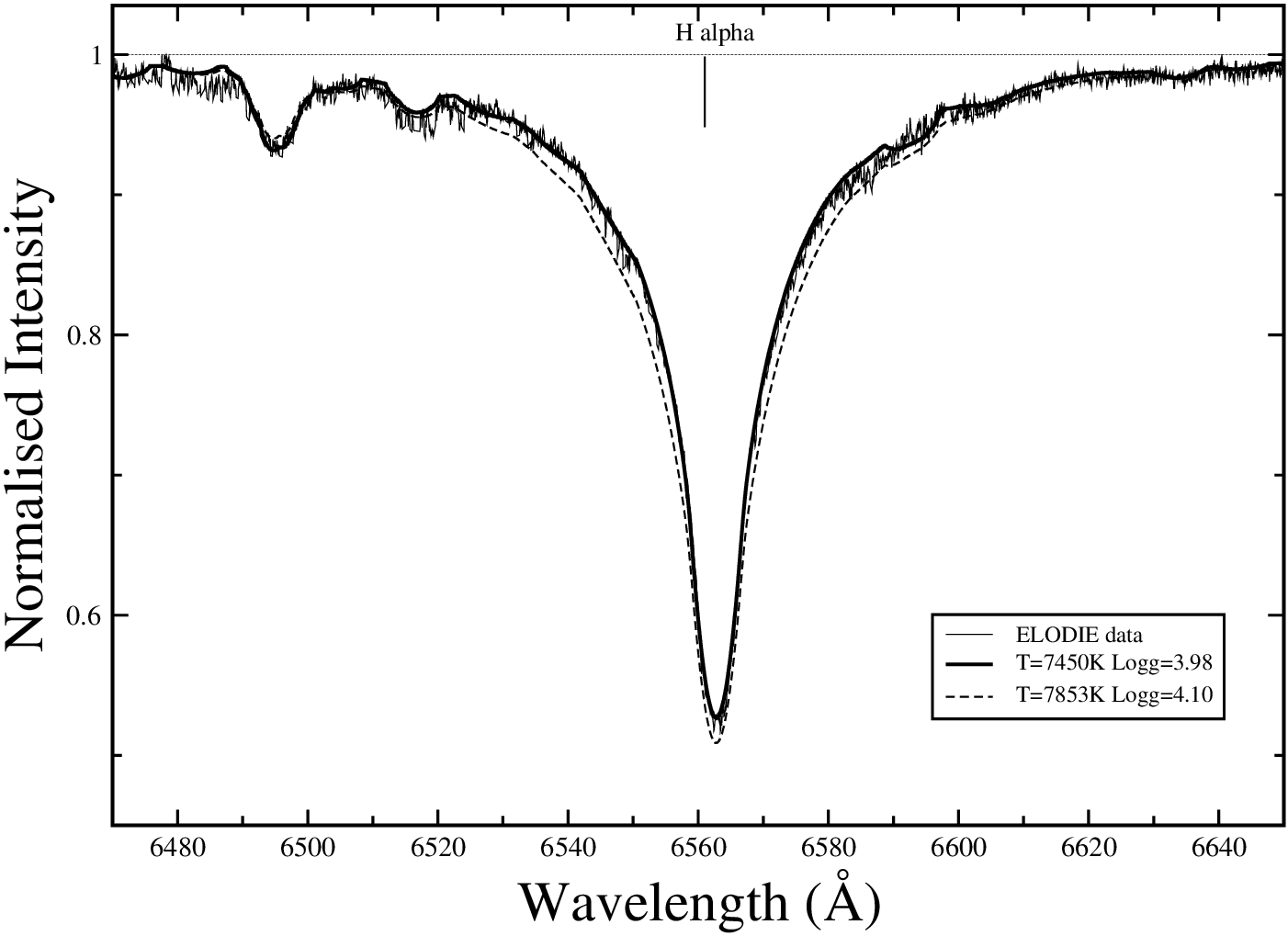}
      \caption{Spectra synthesised using photometric (dashed line) and spectroscopic (solid line) atmosphere parameters, compared to the observed  H$\alpha$ line profile of the ELODIE spectrum.}
\end{center}
 \label{fig1}
\end{figure}

The projected rotational velocities ($v \sin i$) from the ELODIE and HIDES spectra of \astrobj{DD UMa} were calculated as $188\pm14$ and $186\pm11$ km/s, respectively. These values are much higher than the 145 km/s derived by \citet{abtmorrell} who used Gaussian fits to line profiles, but are in excellent agreement with \citet{takeda} who derived 187 km/s by fitting synthetic line profiles to observations. 

Due to the fast rotation of \astrobj{DD UMa}, the spectral lines in our observations are all heavily blended, and thus the equivalent widths of lines can not be measured. Instead, full synthetic spectra were produced with Synth3 \citep{kochukhov}.  Fitting for chemical abundances was done with the Lispan code's line core fitting routine \citep{stutz}, using the blended, rotational broadened synthetic spectra.  For lines blended with other elements, abundances for both elements were determined in an iterative process, relying on additional lines of those elements to provide further constraints.  We adopted the \citet{asplund} solar abundances as the initial point from which to begin the fitting process. Table 2 presents the average abundance values and estimated internal error of each ion, for both the ELODIE and HIDES observations. Due to the fast rotation of DD UMa, the internal error calculated for each ion is relatively large compared to slow rotating stars.

The internal error ($\sigma$) is the standard deviation of the abundances determined for individual lines, and takes into account uncertainties in line oscillator strengths and the normalisation \citep{fossati9}. These values do not take into account the uncertainties in $T_{eff}$, $\log g$, or $\xi$ and lead to an underestimate of the actual abundance uncertainty. In order to derive the actual uncertainty for the abundances determined from both sets of observations, we calculated the iron abundance change for a 1$\sigma$ difference in each fundamental parameter. For example, a $1\sigma$(=150K) change in the $T_{eff}$(=7450 K) is equal to a 0.21 dex Fe abundance change for the ELODIE spectra. Likewise, we calculated the Fe abundance changes due to a difference in $\log g$ and $\xi$ of 1$\sigma$ and present these values in Table 3. When taking all of these uncertainties for Fe into account, the systematic error of the ELODIE and HIDES spectra are 0.28 dex and 0.33 dex, respectively.  

 \begin{table*}

\begin{center}
\caption{The average abundance ($\log(N_X/N_{tot})$) of each ion, internal error ($\sigma$), and number of lines used (\#) for the ELODIE and HIDES spectra of \astrobj{DD UMa}. For comparison, solar abundances from \citet{asplund} are included. }

\begin{tabular}{c@{ }lccc|c@{}cc|c}
\hline\hline
& & ELODIE  & & &   &HIDES\\ 
\hline
At.Number & Ion& $\log(N_X/N_{tot})$& $\sigma$ & \# & $\log(N_X/N_{tot})$&  $\sigma$ & \#  & Solar Abundances\\
 \hline
11 & Na I & -5.36 & 0.14 & 2 & - & - & -&-5.87\\
12 & Mg I & -4.14 & -& 1 & -4.43& -& 1&-4.51\\
12 & Mg II & -4.29 & 0.58 & 3& -& -&- &-4.51\\
14 & Si I & -4.27 & -& 1& -& -&- & -4.53\\
14 & Si II & -4.25 & 0.07& 3& -& -& -& -4.53\\
20 & Ca I & -5.60& 0.39& 10 & -5.76 & 0.01 & 2& -5.73 \\
20 & Ca II & -5.29& -& 1& -& -& -& -5.73\\
21 & Sc II & -9.28& 0.32& 4& -9.00& -& 1& -8.99\\
22 & Ti I & -7.39& 0.24& 3& -7.28& 0.35& 5& -7.14\\
22 & Ti II & -7.36& 0.37& 5& -& -& -& -7.14\\
24 & Cr I & -6.54& 0.16& 5& -6.60 & 0.20& 2& -6.40\\
24 & Cr II & -6.52& 0.18& 2& -6.59& 0.32& 3& -6.40\\
25 & Mn I & -6.50& 0.06& 3& -6.82& 0.14& 2& -6.65\\
26 & Fe I & -4.64& 0.23& 24& -4.71& 0.16& 21 & -4.59\\
26 & Fe II & -4.61& 0.13& 8& -4.69& 0.26& 4 & -4.59\\
28 & Ni I & -6.21& 0.12& 7& -6.00& -& 1& -5.81\\
56 & Ba II & -9.65& 0.45& 3& -& -&-& -9.87 \\

\hline

\end{tabular} 
\end{center}

\label{Table2}
\end{table*}


\begin{figure}[htb]
\begin{center}  
 \includegraphics[trim = 0mm 0mm 0mm 0mm, clip, width=7.8cm]{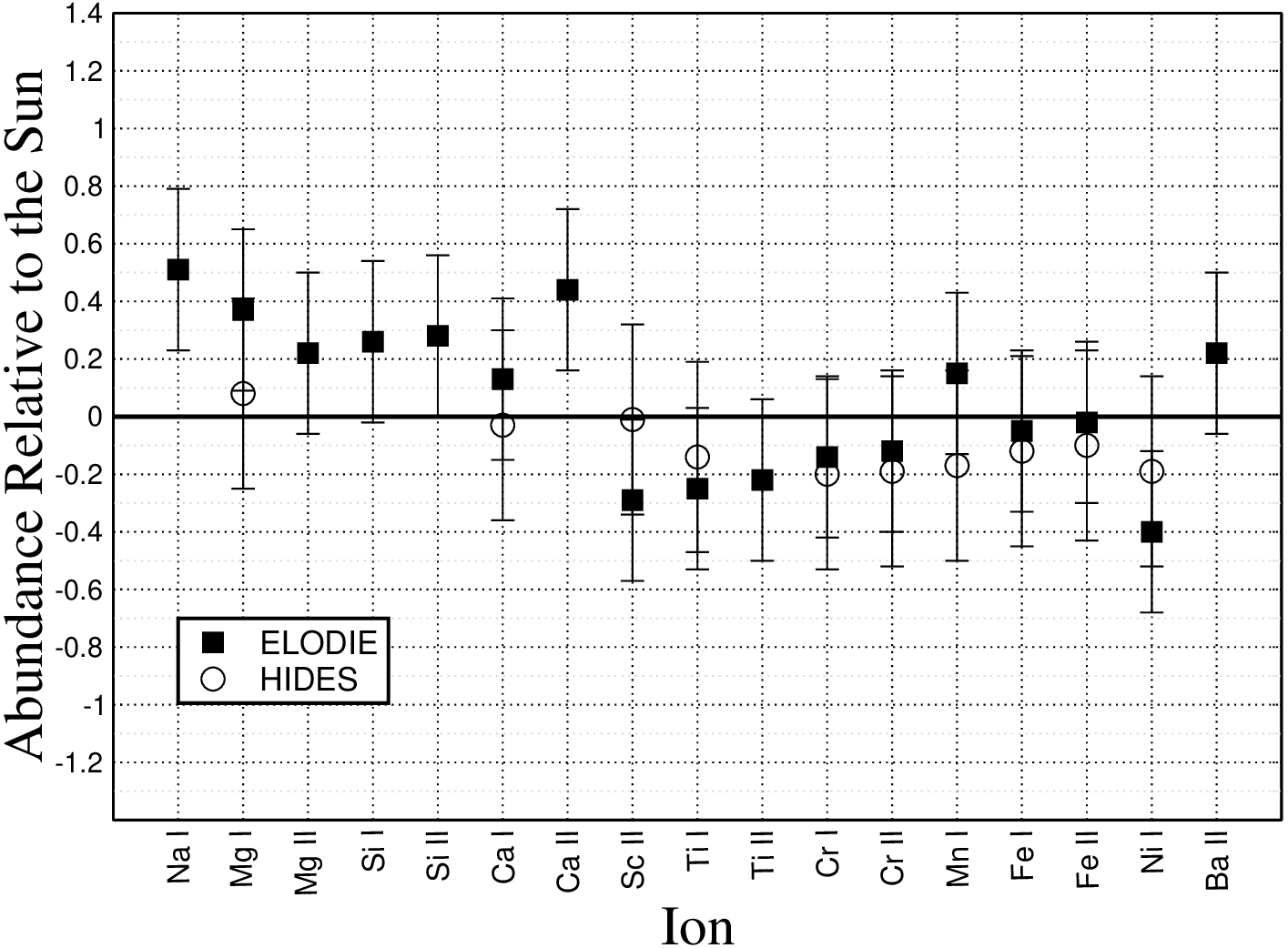}
      \caption{The difference between our derived abundances and the solar abundances of \citet{asplund} for ELODIE (filled square) and HIDES (empty circle) spectra of \astrobj{DD UMa}.  The abundance error bars of ELODIE and HIDES spectra are 0.28 and 0.33 dex respectively, reflecting the systematic uncertainty due to atmospheric parameters. }
\end{center}
 \label{fig2}
\end{figure}

The abundance values obtained from both ELODIE and HIDES spectra are plotted relative to solar abundances \citep{asplund} in Fig. 2. The abundances derived for different ions of the same element are in good agreement for both sets of spectra.  The abundance values of each ion are consistent with the solar values, within the internal and systematic uncertainty, except for Na I and Ca II. Only one line was used to derive the abundance value of Ca II, and thus this value is rather uncertain. The Na abundance was derived from the Na doublet at 5682 \& 5688 \AA\, which is subject to non-LTE effects, and therefore is clearly overestimated. Excluding these two odd ions, \astrobj{DD UMa} has solar abundances with a metallicity of $Z=0.013\pm0.003$. $Z$ was calculated from
\begin{equation}
Z=\frac{\sum_{a\geqslant3} m_a 10^{log(N_a/N_{tot})}}{\sum_{a\geqslant1} m_a 10^{log(N_a/N_{tot})}},
\end{equation}
where $a$ is the atomic number of an element with atomic mass $m_a$ and solar abundances from \citet{asplund} were adopted for all the elements that we did not analyse.
This value is well within uncertainty of the solar value $Z_\odot=0.0122$ reported by \cite{asplund}. 

\begin{table}


\begin{center}
\caption{Changes in the iron abundance for a 1$\sigma$ difference in $T_{eff}$, $\log g$, and $\xi$ for ELODIE and HIDES observations. These represent systematic uncertainties due the different atmospheric parameters. The combined systematic uncertainty is in the last row.} 

\begin{tabular}{ccc}
\hline\hline
  Parameter &ELODIE&HIDES\\
\hline $\sigma_{Teff}$ & 0.21 & 0.23\\
$\sigma_{logg}$ &0.03 &0.03 \\
$\sigma_{\xi} $ &0.18 &0.24 \\
\hline$\sigma_{systematic}$&0.28&0.33\\
 \hline
\end{tabular}

\end{center}

\label{Table3}
\end{table}

\section{Membership and Evolutionary Status}
\label{Membership-Evolutionary}

We assessed the membership of \astrobj{DD UMa} in the \astrobj{Ursa Major group} by means of two criteria: kinematic and spectroscopic. First, we measured the radial velocity from our spectra and calculated the kinematic parameters of \astrobj{DD UMa}. Then we compared these values with the criteria in the literature for the nucleus and extended stream of the \astrobj{Ursa Major group}. Our second approach was to compare the spectroscopic mean iron abundance of the \astrobj{Ursa Major group} \citep{ammler} with the iron abundance of \astrobj{DD UMa}.

We derived the $U$, $V$, and $W$ space motions in Galactic coordinates for \astrobj{DD UMa} with the equations given by \citep{johnson}. 
The parameters used to calculate the space motion are listed in Table 4. 
In this table, the radial velocity was measured from the time-series of HIDES data. 
Parallax and proper motions were adopted from the new reduction of HIPPARCOS data by \citet{vanleeuwen07}. 
The kinematic parameters of the Ursa Major subgroups (nucleus and extended stream) and \astrobj{DD UMa} are listed in Table 5. We compared the kinematic values of \astrobj{DD UMa} with both subgroups and determined that the space motion values of \astrobj{DD UMa} are consistent with the extended stream 
\citep{chereul}, but not with the nucleus \citep{king} when taking into account the uncertainties.  

\begin{table}


\begin{center}
\caption{The parallax ($\pi$) and proper motion ($\mu_\alpha$, $\mu_\delta$) of \astrobj{DD UMa} were adopted from the HIPPARCOS catalogue \citep{vanleeuwen07}, and the radial velocity ($V_R$) was calculated from the HIDES spectra.}

\begin{tabular}{ccc}
\hline\hline  Parameter& Value \\
\hline 
$\pi$ (mas)          & $ 27.55 \pm 0.80$ \\
$\mu_\alpha$ (mas/yr) & $ 49.14 \pm 0.98$ \\
$\mu_\delta$ (mas/yr) & $ 59.79 \pm 0.54$ \\
$V_R$ (km/s)         & $-16.45 \pm 0.93$ \\
 \hline
\end{tabular}

\end{center}

\label{Table4}
\end{table}

\begin{table*}


\begin{center}
\caption{The space velocity in Galactic coordinates ($U$, $V$, and $W$) of the nucleus and extended stream of the \astrobj{Ursa Major group}, and of \astrobj{DD UMa}.}

\begin{tabular}{ccccc}
\hline\hline
  & \multicolumn{3}{c}{space velocity (km/s)} &Reference\\
  & \textbf{$U$}&\textbf{$V$}&\textbf{$W$}&\\
\hline 
UMa Nucleus &$13.9 \pm 0.6$ &$2.9 \pm 0.9$ &$-8.4 \pm 1.3$ & \citet{king}\\
UMa Stream  &$14.0 \pm 7.3$ &$1.0 \pm 6.4$ &$-7.8 \pm 5.5$ & \citet{chereul}\\
\astrobj{DD UMa}      &$18.9 \pm 0.7$ &$6.5 \pm 0.2$ &$-6.7 \pm 0.6$ & This Study \\

 \hline
\end{tabular}

\end{center}

\label{Table5}
\end{table*}

 \citet{ammler} calculated iron abundances, relative to solar, for 17 members of the \astrobj{Ursa Major group} and obtained the average value of [Fe$/$H]$=-0.034\pm0.05$ dex, which is plotted as a horizontal dashed line in Fig. 3. The [Fe$/$H] value of \astrobj{DD UMa} is $-0.04\pm0.20$ dex, based on the ELODIE observation, and is plotted as a filled square in Fig. 3. This figure shows that the iron abundance of \astrobj{DD UMa} is in good agreement with the other members of the \astrobj{Ursa Major group} (plotted as open circles). Furthermore this figure shows that \astrobj{DD UMa} and the \astrobj{Ursa Major group} members have similar compositions.

\begin{figure}[htb]
\begin{center}  
 \includegraphics[trim = 0mm 0mm 0mm 0mm, clip, width=7.8cm]{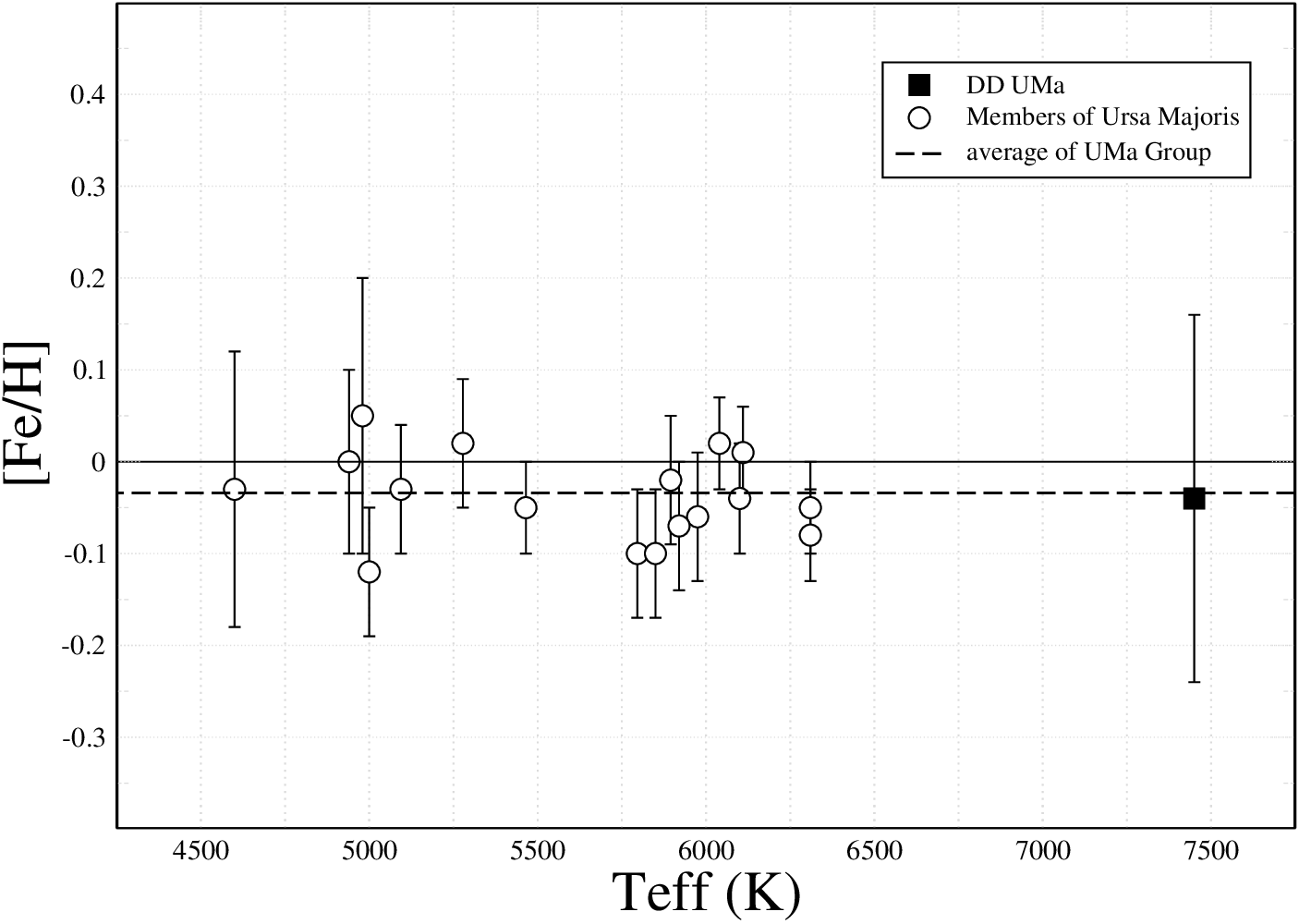}
      \caption{ Comparison of [Fe$/$H] values of the \astrobj{Ursa Major group} members analysed by \citet{ammler} with \astrobj{DD UMa}. The x-axis represents the effective temperature of the stars and the dashed horizontal line is the average [Fe$/$H] value of the \astrobj{Ursa Major group}.}
\end{center}
 \label{fig3}
\end{figure}

\begin{figure}[htb]
\begin{center}  
 \includegraphics[trim = 0mm 0mm 0mm 0mm, clip, width=7.8cm]{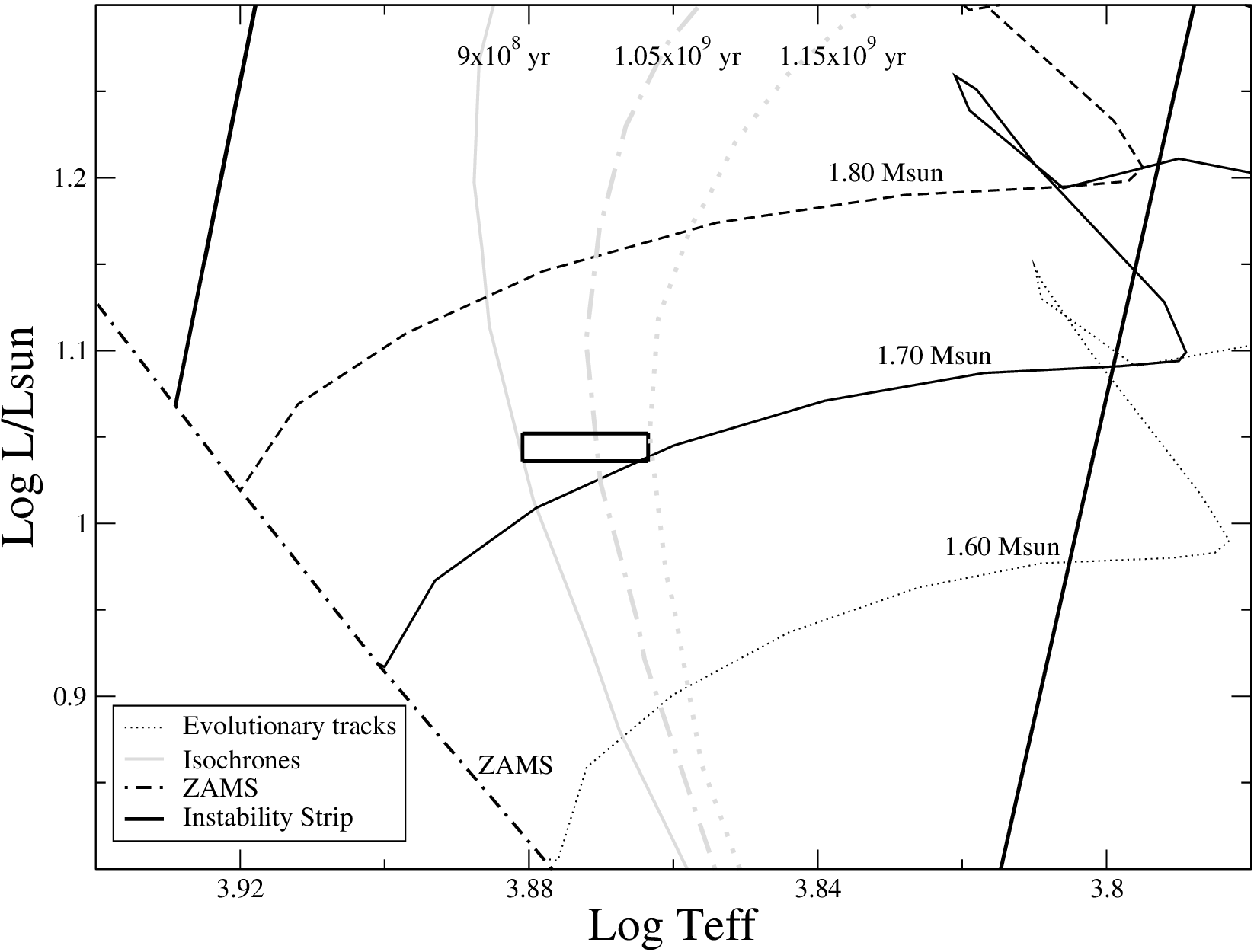}
      \caption{The position of \astrobj{DD UMa} within the instability strip (vertical black lines) on the H-R diagram.  Evolutionary tracks are from \citet{salasnich} and isochrones from \citet{marigo}. The black dotted, full, and dashed lines show evolutionary tracks of 1.60 M$_\odot$, 1.70 M$_\odot$, and 1.80 M$_\odot$ respectively. Isochrones for $9\times10^8$, $1.05\times10^9$, and $1.15\times10^9$ yrs are represented with grey lines from left to right on the H-R diagram.}
      \end{center}
 \label{fig4}
\end{figure}

\begin{table}


\begin{center}
\caption{The apparent magnitude in the V band ($m_V$), parallax $(\pi$), absolute magnitude ($M_V$), bolometric correction, luminosity ($\log(L/L_\odot)$), and logarithmic effective temperature ($\log(T_{eff})$) of \astrobj{DD UMa}. The references give the source of the observation or the calibration used, where applicable. } 

\begin{tabular}{cccc}

\hline\hline
Parameter       & Value             & Reference\\
\hline
$m_V$           & $4.832 \pm 0.012$ & \citet{mermilliod} \\
$\pi$ (mas)     & $27.90 \pm 0.20 $ &  \citet{vanleeuwen07} \\
$M_V$           & $2.060 \pm 0.020$ & \\ 
Bolometric Correction&$0.050\pm0.006$& \citet{balona} \\
$\log(L/L_\odot)$ & $1.044 \pm 0.008$ & \\  
$\log(T_{eff})$   & $3.872 \pm 0.009$ & \\ 
 \hline
\end{tabular}

\end{center}

\label{Table6}
\end{table}

We calculated the luminosity of \astrobj{DD UMa} from the apparent magnitude and parallax, listed in Table 6. 
The $T_{eff}$ used was derived in Sect. \ref{AbundanceAnalysis}.
Using this luminosity and effective temperature we placed \astrobj{DD UMa} on the H-R diagram.  The 1$\sigma$ confidence interval around \astrobj{DD UMa} is denoted by a rectangle in Fig. 4.
\astrobj{DD UMa} is clearly located within the instability strip (vertical black lines, in Fig. 4), which is consistent with it being a $\delta$ Scuti pulsator.

Evolutionary tracks were taken from the computations by \citet{salasnich} and are plotted in the H-R diagram as black lines for masses of 1.60 M$_\odot$ (dotted), 1.70 M$_\odot$ (dashed), and 1.80 M$_\odot$ (full), with solar metallicity. 
By comparison to these evolutionary tracks, we conclude that \astrobj{DD UMa} has a mass of $1.72 \pm0.02$ M$_\odot$. 
Figure 4 shows three isochrones with solar metallicity from \citet{marigo} for ages of $9\times10^8$, $1.05\times10^9$, and $1.15\times10^9$ years, from left to right. 
The \astrobj{Ursa Major group} contains stars loosely clustered around the ages $\sim$$10^7$, $6\times10^8$, and $1.5\times10^9$ years \citep{chereul}. 
\astrobj{DD UMa} has a main sequence age of $1.05^{+0.10}_{-0.15} \times10^9$ years, which is well within the age range of group members, and on the older side.

\section{Conclusions}

We conclude that kinematically \astrobj{DD UMa} is located in the extended stream of the \astrobj{Ursa Major group}, 
and that it satisfies the spectroscopic criteria given by \citet{ammler} for a group member. 

\astrobj{DD UMa} has solar chemical abundances, an effective temperature of $7450\pm150$ K  and a $\log g$ of $3.98\pm0.2$.  From the star's H-R diagram position we infer a mass of $1.72 \pm0.02$ M$_\odot$ and an age of $1.05^{+0.10}_{-0.15} \times10^9$ years, which makes it coeval with the \astrobj{Ursa Major group}.  As a $\delta$ Scuti pulsator, we identified a main pulsational frequency of 9.4 c/d, and a second frequency of 15.0 c/d.  

\astrobj{DD UMa} is an unbound and relatively old (compared to the group average) member of the \astrobj{Ursa Major group}. It will provide important information for subsequent kinematic evolution and origin studies of the \astrobj{Ursa Major group}. Furthermore, this study provides valuable information for future asteroseismic studies with the BRITE-constellation, which is planned to launch this year. Photometric measurements with the BRITE-constellation will reveal more pulsation frequencies for mode identification, which can be compared to detailed asteroseismic models.




\bigskip
{\bf \noindent Acknowledgements}
 This paper is based on observations using the HIDES spectrograph at the Okayama Astrophysical Observatory (Japan). A. Elmasl\i\, gratefully acknowledges the National Astronomical Observatory of Japan (NAOJ) for providing financial support for the observing run (Project no: 09A-12). She also acknowledges Ankara University for providing her a travel grant to the UK and the Open University (UK) for funding accommodation during the analysis of the data.

\bibliographystyle{model2-names}
\bibliography{dduma} 









\end{document}